\documentclass[11pt,a4paper]{article}
\usepackage[T1]{fontenc}
\usepackage[utf8]{inputenc}
\usepackage{authblk}
\usepackage{graphicx}
\usepackage{amsmath,amsfonts}
 \begin{document}
\title{Constructing Quantum Logic Gates Using q-Deformed Harmonic Oscillator Algebras}
\author[1]{A. A. Altintas\footnote{e-mail:\;altintas.azmiali@gmail.com}
}
\author[2]{F. Ozaydin}
\author[3]{C. Yesilyurt}
\author[4]{S. Bugu}
\author[5]{M. Arik}
\affil[1]{Department of Electrical Engineering, Faculty of Engineering and Architecture, Okan University, Istanbul, Turkey}
\affil[2]{Department of Information Technologies, Faculty of Science and Letters, Isik University, Istanbul, Turkey}
\affil[3]{Department of Computer Engineering, Faculty of Engineering and Architecture, Okan University, Istanbul, Turkey}
\affil[4]{Department of Computer Engineering, Institute of Science, Istanbul University, Istanbul, Turkey}
\affil[5]{Physics Department, Bogazici University,  Istanbul, Turkey}

\maketitle
\begin{abstract}
We study two-level q-deformed angular momentum states and using q-deformed harmonic oscillators,
we provide a framework for constructing qubits and quantum gates. We also present the construction of some basic quantum logic gates, including CNOT, SWAP, Toffoli and Fredkin.
\end{abstract}
\section{Introduction}
Solving the quantum integrable models, quantum groups have been studied since the eighties. Among many, one of the basic methods to construct quantum groups is the Drinfeld's method \cite{drinfeld}, in which a deformation parameter, such as $q$, is defined on usual Lie algebra. Woronowicz's approach \cite{wor} involves non-commutative co-multiplication and in Manin's approach \cite{manin} one can construct a quantum group by making linear transformations on quantum plane. Recently Manin's approach is used to find inhomogeneous invariance quantum groups of various particle algebras \cite{ali,ali2,ali3,ali4,ali5,ali6,ali7,ali8,ali9,ali10}.

The particle algebras play an important role to describe fermions and bosons both in quantum mechanics and quantum field theory. One can define deformation parameter on particle algebra instead of Lie algebra. Then the new system is called as deformed particle algebra. The first examples of deformed particle algebras are about generalization of boson algebra \cite{arik,arik2} by using a real deformation parameter $q$. Macfarlane \cite{macfarlane} and Biedenharn \cite{biedenharn} showed the relation between $q$ deformed boson algebra and $SU_q(2)$, which means $q$ deformed special unitary group in two dimensions.
Recently the q-deformed particle algebras is used to understand the difference between theoretical and experimental results in nuclear and atomic physics \cite{bonatsos,bonatsos2,Georgieva}.

Recently entanglement measures of bipartite systems has been studied in the context of q-deformed Heisenberg-Weyl algebra and a large class of bipartite entangled states are obtained. The formalism of entanglement is generalized to mixed states\cite{Berrada}.  Bipartite entanglement in the context of $U_q(su(2))$ quantum algebra used spin coherent state is introduced and these entangled states has richer structure than non-deformed case\cite{Berrada2}.

Also it was shown that one can give good explanation to semiconductor-cavity QED in high Q- regimes constructing excitons by using q deformed boson algebra and it fits experiments\cite{Sun}. q-deformed harmonic oscillators, on the other hand, can provide models for quantum computations with such as ion traps \cite{Sharma} and q-deformed relative entropy has been shown to be playing an important role in the extensivity of quantum information metrics \cite{Hasegawa}.

Since qubits can be expressed using angular momentum states $|jm\rangle$, natural questions arise: \textit{i)} Using q-deformed angular momentum states, can one construct qubits; \textit{ii)} How do the quantum gates act on q-deformed qubits; and \textit{iii)} Can one construct q-deformed quantum gates. In the following sections, we present our results for these questions.

\section{Qubits and Gates}

In this section we introduce qubits in terms of angular momentum states and formulate the operation of quantum logic gate to any qubit.
Schwinger representation \cite{SchwingerBook} of angular momentum states are written as
\begin{equation}
|jm\rangle=\frac{(a_1^\dag)^{j+m} (a_2^\dag)^{j-m}}{\sqrt{(j+m)!(j-m)!}}|\tilde{0_1}\tilde{0_2}\rangle.
\end{equation}
Here $j$ is the total angular momentum quantum number and $m$ is the $z^{th}$ component of $j$, then $a_1^{\dag}$ is the creation operator for $j$ and
$a_2^{\dag}$ is the creation operator for $m$. Also $|\tilde{0_1}\rangle$ is oscillator ground state of $j$ and $|\tilde{0_2}\rangle$ is the oscillator ground state of $m$.
Substituting $j+m=n_1$, $j-m=n_2$ to Eq.(1),

\begin{equation}
|jm\rangle=\frac{(a_1^\dag)^{n_1} (a_2^\dag)^{n_2}}{\sqrt{(n_1)!(n_2)!}}|\tilde{0_1}\tilde{0_2}\rangle,
\end{equation}

where $n_1+n_2=1$ and $n_1,n_2=\{0,1\}$. $n_1$ and $n_2$ are the eigenvalues of the number operators.
The creation and annihilation operators  satisfy following relations
 \begin{equation}
aa^\dag-a^\dag a=1,
 \end{equation}
\begin{equation}
[a,N]=a,\;\;\; [a^\dag, N]=-a^\dag,
\end{equation}
\begin{equation}
N=a^\dag a,\;\;\; N^\dag= N.
\end{equation}

To construct a qubit we must define the values of $j$ and $m$. In order to write a qubit by using angular momentum state $j$ must be equal to $1/2$. Then $m$ can take values $1/2$ and $-1/2$, which means that for $j=1/2$
there are two possible states $|1/2\; 1/2\rangle$ and $|1/2\; -1/2\rangle$, corresponding to qubit states,
\begin{eqnarray}
|1/2\;-1/2\rangle&=|0\rangle,\\
|1/2\;1/2\rangle&=|1\rangle.
\end{eqnarray}
Since $j$'s are the same for both states only $m$ is the determining parameter. As a short hand notation we can express spin down state as $|-1/2\rangle$ and spin up state
as $|1/2\rangle$.

Using Eq.(1) we can write qubit states in terms of oscillator states,
\begin{eqnarray}
|0\rangle&=a_2^{\dag}|\tilde{0_1}\tilde{0_2}\rangle&=|\tilde{0_1}\tilde{1_2}\rangle,\\
|1\rangle&=a_1^{\dag}|\tilde{0_1}\tilde{0_2}\rangle&=|\tilde{1_1}\tilde{0_2}\rangle.
\end{eqnarray}
Taking $j=1/2$ in $j+m=n_1$ one can write $m$ as $n_1-1/2$ then  a general formula for qubit states can be written as;
\begin{equation}
|n_1-1/2\rangle=\frac{(a_1^\dag)^{n_1} (a_2^\dag)^{(n_2)}}{\sqrt{(n_1)!(n_2)!}}|\tilde{0_1}\tilde{0_2}\rangle
\end{equation}
With the help of Eq.(10) any qubit state can be written. For spin up state $n_1$ must be $1$ and for spin down state $n_1$ must be $0$. Thus
Eq.(10) is simplified to
\begin{equation}
|x\rangle=(a_1^\dag)^{x} (a_2^\dag)^{(1-x)}|\tilde{0_1}\tilde{0_2}\rangle,
\end{equation}
where $x$ can  have values $0$ and $1$.
In addition to that if one uses $j-m=n_2$ to write an expression for $m$, the expression will be $1/2-n_1$. Thus a qubit state can also be written as below
\begin{equation}
|1/2-n_1\rangle=\frac{(a_1^\dag)^{n_2} (a_2^\dag)^{(n_1)}}{\sqrt{(n_1)!(n_2)!}}|\tilde{0_1}\tilde{0_2}\rangle.
\end{equation}
 Eq.(12) is used to write a qubit state $|1-x\rangle$. Substituting $x$ for $n_1$ one can get
\begin{equation}
|1-x\rangle=(a_1^\dag)^{1-x} (a_2^\dag)^{(x)}|\tilde{0_1}\tilde{0_2}\rangle.
\end{equation}

Recall that $n_1$ and $x$ can take values only $0$ and $1$.

Let us now express how the gates act on qubits which are written in terms of oscillator states.  The operation of Phase Shift (PS), Hadamard (Had), Not and Controlled-Not (CNot) gates \cite{RefNC} has been given in \cite{Gangopad2} as
\begin{eqnarray}
\mathbf{PS}|x\rangle&=&e^{ix\phi}|x\rangle,\\
\mathbf{Had}|x\rangle&=&(-1)^x|x\rangle+|1-x\rangle,\\
\mathbf{Not}|x\rangle&=&|1-x\rangle,\\
\mathbf{CNot}|x\;y\rangle&=&(1-x)|x\;y\rangle+x|x\;1-y\rangle.
\end{eqnarray}

Note that in Eqs.(14-16) $|x\rangle$ is an single qubit state but in equation (17) $|x\;y\rangle$ is a two qubit state which is expressed as $|x\;y\rangle=|x\rangle|y\rangle$.

Next, we will write Swap, Toffoli (control-control-not) and Fredkin (control-swap) gates. Note that Toffoli and Fredkin \cite{ToffoliFredkin1982} gates are essential for many quantum information tasks, and recently there is an increasing interest on these gates for creating large-scale W state quantum networks \cite{BuguPRA2013A,YesilyurtQInP2013A,Ozaydin2014PRA}.

\begin{eqnarray}
\mathbf{Swap}|x\;y\rangle&=&|y\;x\rangle,\\
\mathbf{Fredkin}|x\;y\;z\rangle&=&(1-x)|x\;y\;z\rangle+x|x\;z\;y\rangle,\\
\mathbf{Toffoli}|x\;y\;z\rangle&=&(xy)|x\;y\;1-z\rangle+[(1-x)y+(1-y)x\\
&+&(1-y)(1-x)]|x\;y\;z\rangle.\nonumber
\end{eqnarray}

\section{q-Deformed Oscillator and q-deformed Qubits}

In this section we rewrite equations $(14-20)$ by using q-deformed harmonic oscillators. To do that we must define a q-deformed oscillator algebra.
 Following equations $(3-5)$ we write q-deformed boson algebra. The q-deformed boson algebra can be written as in \cite{lohe}
\begin{equation}
a_qa_q^\dag-qa_q^\dag a_q=q^{-N},\\
\end{equation}
\begin{equation}
[a_q,N_q]=a_q,\;\;\; [a_q^\dag, N_q]=-a_q^\dag
\end{equation}
\begin{equation}
[N]_q=a_q^\dag a_q,\;\;\;a_qa_q^\dag=[N+1]_q.
\end{equation}
Here $a_q$ and $a_q^\dag$ are q deformed annihilation and creation operators respectively. Unlike the usual boson algebra,  the operator $a_q^\dag a_q$ is not a number operator but it is equal to $[N]_q$ and a deformed  number can be defined as $[n]=\frac{q^n-q^{-n}}{q-q^{-1}}.$ Here q is a deformation parameter $q=e^s$ and $q \in$ $\mathbf{R^+}$.

For a general realization the relation between q-deformed operators and usual operators is given in \cite{Gangopad}
\begin{eqnarray}
a_q&=&\sqrt{\frac{q^N\psi_1-q^{-N}\psi_2}{N(q-q^{-1})}}a,\\
a_q^\dag&=&\sqrt{\frac{q^N\psi_1-q^{-N}\psi_2}{N(q-q^{-1})}}a^\dag,\\
N_q&=&N-(1/s)ln\psi_2.
\end{eqnarray}
Here $\psi_1$ and $\psi_2$ are arbitrary parameters which depends on deformation parameter $q$. When they equal to 1 one can reach non-deformed boson algebra.
The qubit states $|x\rangle_q$ and $|1-x\rangle_q$ are written by q-deformed oscillators.
\begin{eqnarray}
|x\rangle_q&=&(a_1^\dag)_q^{x} (a_2^\dag)_q^{(1-x)}|\tilde{0_1}\tilde{0_2}\rangle,\\
|1-x\rangle_q&=&(a_1^\dag)_q^{1-x} (a_2^\dag)_q^{(x)}|\tilde{0_1}\tilde{0_2}\rangle.
\end{eqnarray}
By using equations (24) and (25) one can rewrite equations (27) and (28).
\begin{eqnarray}
&|x\rangle_q=\left(\sqrt{\frac{q^{N_1}\psi_1-q^{-N_1}\psi_2}{N_1(q-q^{-1})}}a_1^\dag\right)^x
\left(\sqrt{\frac{q^{N_2}\psi_3-q^{-N_2}\psi_4}{N_2(q-q^{-1})}}a_2^\dag\right)^{1-x}|\tilde{0_1}\tilde{0_2}\rangle,\\
&|1-x\rangle_q=\left(\sqrt{\frac{q^{N_1}\psi_1-q^{-N_1}\psi_2}{N_1(q-q^{-1})}}a_1^\dag\right)^{1-x}
\left(\sqrt{\frac{q^{N_2}\psi_3-q^{-N_2}\psi_4}{N_2(q-q^{-1})}}a_2^\dag\right)^{x}|\tilde{0_1}\tilde{0_2}\rangle
\end{eqnarray}
In order to write equations in  q-deformed case we must replace $|x\rangle$ and $|1-x\rangle$ by  $|x\rangle_q$ and $|1-x\rangle_q$ respectively in equations (14-20). To satisfy equations (14-20) in q deformed case the arbitrary parameters should be defined using equations (29) and (30).

First we use equation (29) in q-deformed version of equation (14) ($\mathbf{PS}|x\rangle_q=e^{ix\phi}|x\rangle_q$). As it can be easily seen from equality there is no restriction on parameters $\psi_1$ and $\psi_2$.

For the Hadamard gate ($\mathbf{Had}|x\rangle_q=(-1)^x|x\rangle_q+|1-x\rangle_q$). By using equations (29) and (30) one can find $\psi_1=\psi_2$ by assuming $\psi_1=\psi_3$ and $\psi_2=\psi_4$.

For the Not gate ($\mathbf{Not}|x\rangle=|1-x\rangle$) we have same restrictions as Hadamard case, which means $\psi_1=\psi_2$ by assuming $\psi_1=\psi_3$ and $\psi_2=\psi_4$.

For CNOT gate ($\mathbf{CNot}|x\;y\rangle_q=(1-x)|x\;y\rangle_q+x|x\;1-y\rangle_q$) the parameters must be $\psi_5=\psi_6$ with assumptions $\psi_5=\psi_7$ and $\psi_6=\psi_8$.

For Swap gate ($\mathbf{Swap}|x\;y\rangle_q=|y\;x\rangle_q$), there is no restriction on parameters $\psi_i$, i runs from 1 to 8.

For Fredkin gate ($\mathbf{Fredkin}|x\;y\;z\rangle_q=(1-x)|x\;y\;z\rangle_q+x|x\;z\;y\rangle_q$), there is no restriction on parameters $\psi_i$, i runs from 1 to 12.

For Toffoli gate ($\mathbf{Toffoli}|x\;y\;z\rangle_q=(xy)|x\;y\;1-z\rangle_q+[(1-x)y+(1-y)x+(1-y)(1-x)]|x\;y\;z\rangle_q$), the parameters are $\psi_9=\psi_{10}$ by taking $\psi_9=\psi_{11}$ and $\psi_{10}=\psi_{12}$.

As an example of finding parameters look at Hadamard gate.
\begin{equation}
\mathbf{Had}|x\rangle_q=(-1)^x|x\rangle_q+|1-x\rangle_q
\end{equation}
\begin{eqnarray}
&\mathbf{Had}\left(\sqrt{\frac{q^{N_1}\psi_1-q^{-N_1}\psi_2}{N_1(q-q^{-1})}}a_1^\dag\right)^x
\left(\sqrt{\frac{q^{N_2}\psi_3-q^{-N_2}\psi_4}{N_2(q-q^{-1})}}a_2^\dag\right)^{1-x}|\tilde{0_1}\tilde{0_2}\rangle=\\\nonumber
&(-1)^x\left(\sqrt{\frac{q^{N_1}\psi_1-q^{-N_1}\psi_2}{N_1(q-q^{-1})}}a_1^\dag\right)^x
\left(\sqrt{\frac{q^{N_2}\psi_3-q^{-N_2}\psi_4}{N_2(q-q^{-1})}}a_2^\dag\right)^{1-x}|\tilde{0_1}\tilde{0_2}\rangle+\\\nonumber
&\left(\sqrt{\frac{q^{N_1}\psi_1-q^{-N_1}\psi_2}{N_1(q-q^{-1})}}a_1^\dag\right)^{1-x}
\left(\sqrt{\frac{q^{N_2}\psi_3-q^{-N_2}\psi_4}{N_2(q-q^{-1})}}a_2^\dag\right)^{x}|\tilde{0_1}\tilde{0_2}\rangle.
\end{eqnarray}
Then by arranging the terms in equation 32 one can get
\begin{eqnarray}
&\mathbf{Had}(a_1^\dag)^x(a_2^\dag)^{1-x}|\tilde{0_1}\tilde{0_2}\rangle=(-1)^x(a_1^\dag)^x(a_2^\dag)^{1-x}|\tilde{0_1}\tilde{0_2}\rangle+\\\nonumber
&\frac{\left(\sqrt{\frac{q^{N_1}\psi_1-q^{-N_1}\psi_2}{N_1(q-q^{-1})}}\right)^{1-x}
\left(\sqrt{\frac{q^{N_2}\psi_3-q^{-N_2}\psi_4}{N_2(q-q^{-1})}}\right)^{x}}
{\left(\sqrt{\frac{q^{N_1}\psi_1-q^{-N_1}\psi_2}{N_1(q-q^{-1})}}\right)^x
\left(\sqrt{\frac{q^{N_2}\psi_3-q^{-N_2}\psi_4}{N_2(q-q^{-1})}}\right)^{1-x}}(a_1^\dag)^{1-x}(a_2^\dag)^x|\tilde{0_1}\tilde{0_2}\rangle.
\end{eqnarray}
Then by taking $\psi_1=\psi_3$ and $\psi_2=\psi_4$ one can get by remembering $N_2=1-N_1$
\begin{equation}
\mathbf{Had}(a_1^\dag)^x(a_2^\dag)^{1-x}|\tilde{0_1}\tilde{0_2}\rangle=(-1)^x(a_1^\dag)^x(a_2^\dag)^{1-x}|\tilde{0_1}\tilde{0_2}\rangle+
\left(\frac{\left(\frac{q^{N_1}\psi_1-q^{-N_1}\psi_2}{N_1(q-q^{-1})}\right)}
{\left(\frac{q^{(1-N_1)}\psi_1-q^{(1-N_1)}\psi_2}{-(1-N_1)(q-q^{-1})}\right)}\right)^{\frac{1}{2}-x}(a_1^\dag)^{1-x}(a_2^\dag)^x|\tilde{0_1}\tilde{0_2}\rangle.
\end{equation}
The term in the parenthesis must be 1. By writing the eigenvalues of $N_1$ as $n_1$ one can see that
\begin{equation}
\frac{\psi_1}{\psi_2}=\frac{(1-n_1)q^{-n_1}-n_1q^{n_1+1}}{(1-n_1)q^{n_1}-n_1q^{n_1-1}}
\end{equation}
The result is always 1 since $n_1$ can only takes values 0 and 1. This means that $\psi_1=\psi_2$ for the Hadamard gate.
For the other gates one can do the similar calculations and reach the results which are given above.

section{q-deformed Quantum Logic Gates}

Here we introduce quantum logic gates in terms of q deformed oscillator states. They can be expressed as

\begin{eqnarray}
\mathbf{Not_q}&=&\sum_{x=0}^{1}|1-x\rangle_q\;_q\langle x|,\\
\mathbf{Had_q}&=&-1^{(N_1)}+\sum_{x=0}^{1}|1-x\rangle_q\;_q\langle x|\\
\mathbf{Swap_q}&=&\sum_{x,y=0}^{1}|y\;x\rangle_q\;_q\langle y\;x|,\\
\mathbf{Cnot_q}&=&(1-N_1)+\sum_{x,y=0}^{1}|x\;1-y\rangle_q\;_q\langle y\;x|N_1\\
\mathbf{Fredkin_q}&=&(1-N_1)+\sum_{x,y,z=0}^{1}|x\;z\;y\rangle_q\;_q\langle z\;y\;x|N_1,\\
\mathbf{Toffoli_q}&=&\sum_{x,y,z=0}^{1}|x\;y\;1-z\rangle_q\;_q\langle z\;y\;x|\left[N_1M_1+(1-N_1)M_1\right]\\\nonumber
&+&\sum_{x,y,z=0}^{1}|x\;y\;1-z\rangle_q\;_q\langle z\;y\;x|\left[(1-N_1)M_1+(1-N_1)(1-M_1)\right].
\end{eqnarray}
Here $N_1$ and $M_1$ are number operators for states $|x\rangle$ and $|y\rangle$.
The next step is obvious, to define arbitrary parameter $\psi$'s for acting the gates on q-deformed qubits. In order to do that we take the simplest
case which is q-deformed NOT gate acts q deformed qubit ($\mathbf{Not_q}|x\rangle_q=|1-x\rangle_q$).
\begin{equation}
\mathbf{Not_q}|x\rangle_q=|1-x\rangle_q\;_q\langle x|x\rangle_q
\end{equation}
By using equations (29) and  (30) one can write the right hand side of equation 42 as,

\begin{eqnarray}
\left(\frac{q^{N_1}\psi_1-q^{-N_1}\psi_2}{N_1(q-q^{-1})}\right)^\frac{1-x}{2}(a_1^\dag)^{1-x}
\left(\frac{q^{N_2}\psi_3-q^{-N_2}\psi_4}{N_2(q-q^{-1})}\right)^\frac{x}{2}(a_2^\dag)^{x}|\tilde{0_1}\tilde{0_2}\rangle\\\nonumber
\langle\tilde{0_1}\tilde{0_2}|(a_2)^{1-x}\left(\frac{q^{N_2}\psi_3-q^{-N_2}\psi_4}{N_2(q-q^{-1})}\right)^\frac{1-x}{2}
(a_1)^{x}\left(\frac{q^{N_1}\psi_1-q^{-N_1}\psi_2}{N_1(q-q^{-1})}\right)^\frac{x}{2}\\\nonumber
\left(\frac{q^{N_1}\psi_1-q^{-N_1}\psi_2}{N_1(q-q^{-1})}\right)^\frac{x}{2}(a_1^\dag)^{x}
\left(\frac{q^{N_2}\psi_3-q^{-N_2}\psi_4}{N_2(q-q^{-1})}\right)^\frac{1-x}{2}(a_2^\dag)^{1-x}|\tilde{0_1}\tilde{0_2}\rangle
\end{eqnarray}

With the help of property $af(N)=f(N+1)a$ and by choosing $\psi_3=\psi_4\;, \psi_1=\psi_2$, the parameters can be found as
\begin{equation}
\psi_3=q^{n_1}\;\mbox{and}\;\psi_1=q^{1-n_1}.
\end{equation}

Then finally we can construct q-deformed qubits and gates by using equation 44. The q-deformed qubits are
\begin{eqnarray}
|x\rangle_q&=&(q^{1-n_1})^\frac{x}{2}(a_1^\dag)^{x}(q^{n_1})^\frac{1-x}{2}(a_2^\dag)^{1-x}|\tilde{0_1}\tilde{0_2}\rangle,\\
|1-x\rangle_q&=&(q^{1-n_1})^\frac{1-x}{2}(a_1^\dag)^{1-x}(q^{n_1})^\frac{x}{2}(a_2^\dag)^{x}|\tilde{0_1}\tilde{0_2}\rangle.
\end{eqnarray}

Then the q-deformed gates are
\begin{eqnarray}
\mathbf{Not_q}&=&(q^{1-n_1})^{(\frac{1-x}{2})}(a_1^\dag)^{1-x}(q^{n_1})^{\frac{x}{2}}(a_2^\dag)^x|\tilde{0_1}\tilde{0_2}\rangle
\langle\tilde{0_1}\tilde{0_2}|\\\nonumber
&&(q^{n_1})^{(\frac{1-x}{2})}(a_2)^{1-x}(q^{1-n_1})^{\frac{x}{2}}(a_1)^x,\\
\vspace{2cm}
\mathbf{Had_q}&=-&1^{N_1}+(q^{1-n_1})^{(\frac{1-x}{2})}(a_1^\dag)^{1-x}(q^n_1)^{\frac{x}{2}}(a_2^\dag)^x|\tilde{0_1}\tilde{0_2}\rangle
\langle\tilde{0_1}\tilde{0_2}|(q^{n_1})^{(\frac{1-x}{2})}\\\nonumber
&&(a_2)^{1-x}(q^{1-n_1})^{\frac{x}{2}}(a_1)^x,\\
\mathbf{Swap_q}&=&(q^{1-m_1})^{(\frac{1-y}{2})}(b_1^\dag)^{1-y}(q^m_1)^{\frac{y}{2}}(b_2^\dag)^y|\tilde{0_1}\tilde{0_2}\rangle
(q^{1-n_1})^{(\frac{1-x}{2})}(a_1^\dag)^{1-x}\\\nonumber
&&(q^n_1)^{\frac{x}{2}}(a_2^\dag)^x|\tilde{0_1}\tilde{0_2}\rangle
\langle\tilde{0_1}\tilde{0_2}|(q^{m_1})^{(\frac{1-y}{2})}(b_2)^{1-y}(q^{1-m_1})^{\frac{y}{2}}(b_1)^y
\langle\tilde{0_1}\tilde{0_2}|\\\nonumber
&&(q^{n_1})^{(\frac{1-x}{2})}(a_2)^{1-x}(q^{1-n_1})^{\frac{x}{2}}(a_1)^x,\\
\mathbf{Cnot_q}&=&(1-N_1)+(q^{1-n_1})^\frac{x}{2}(a_1^\dag)^{x}(q^{n_1})^\frac{1-x}{2}(a_2^\dag)^{1-x}|\tilde{0_1}\tilde{0_2}\rangle
(q^{1-m_1})^\frac{1-y}{2}(b_1^\dag)^{1-y}\\\nonumber
&&(q^{m_1})^\frac{y}{2}(b_2^\dag)^{y}|\tilde{0_1}\tilde{0_2}\rangle
\langle\tilde{0_1}\tilde{0_2}|(q^{m_1})^{(\frac{1-y}{2})}(b_2)^{1-y}(q^{1-m_1})^{\frac{y}{2}}\\\nonumber
&&(b_1)^y\langle\tilde{0_1}\tilde{0_2}|(q^{n_1})^{(\frac{1-x}{2})}(a_2)^{1-x}(q^{1-n_1})^{\frac{x}{2}}(a_1)^x(N_1),\\
\mathbf{Fredkin_q}&=&(1-N_1)+(q^{1-n_1})^\frac{x}{2}(a_1^\dag)^{x}(q^{n_1})^\frac{1-x}{2}(a_2^\dag)^{1-x}|\tilde{0_1}\tilde{0_2}\rangle
(q^{1-p_1})^\frac{z}{2}(c_1^\dag)^{z}\\\nonumber
&&(q^{p_1})^\frac{1-z}{2}(c_2^\dag)^{1-z}|\tilde{0_1}\tilde{0_2}\rangle
(q^{1-m_1})^\frac{y}{2}(b_1^\dag)^{y}(q^{m_1})^\frac{1-y}{2}(b_2^\dag)^{1-y}|\tilde{0_1}\tilde{0_2}\rangle
\langle\tilde{0_1}\tilde{0_2}|\\\nonumber
&&(q^{p_1})^{(\frac{1-z}{2})}(c_2)^{1-z}(q^{1-p_1})^{\frac{z}{2}}(c_1)^z
\langle\tilde{0_1}\tilde{0_2}|(q^{m_1})^{(\frac{1-y}{2})}(b_2)^{1-y}\\\nonumber
&&(q^{1-m_1})^{\frac{y}{2}}(b_1)^y
\langle\tilde{0_1}\tilde{0_2}|(q^{n_1})^{(\frac{1-x}{2})}(a_2)^{1-x}(q^{1-n_1})^{\frac{x}{2}}(a_1)^x,\\
\mathbf{Toffoli_q}&=&(q^{1-n_1})^{(\frac{1-x}{2})}(a_1^\dag)^{1-x}(q^n_1)^{\frac{x}{2}}(a_2^\dag)^x|\tilde{0_1}\tilde{0_2}\rangle
(q^{1-m_1})^{(\frac{1-y}{2})}(b_1^\dag)^{1-y}(q^m_1)^{\frac{y}{2}}\\\nonumber
&&(b_2^\dag)^y|\tilde{0_1}\tilde{0_2}\rangle(q^{1-p_1})^\frac{1-z}{2}(c_1^\dag)^{1-z}(q^{p_1})^\frac{z}{2}(c_2^\dag)^{z}|\tilde{0_1}\tilde{0_2}\rangle
\langle\tilde{0_1}\tilde{0_2}|(q^{p_1})^{(\frac{1-z}{2})}(c_2)^{1-z}(q^{1-p_1})^{\frac{z}{2}}(c_1)^z\\\nonumber
&&\langle\tilde{0_1}\tilde{0_2}|(q^{m_1})^{(\frac{1-y}{2})}(b_2)^{1-y}(q^{1-m_1})^{\frac{y}{2}}(b_1)^y\\\nonumber
&&\langle\tilde{0_1}\tilde{0_2}|(q^{n_1})^{(\frac{1-x}{2})}(a_2)^{1-x}(q^{1-n_1})^{\frac{x}{2}}(a_1)^x[(N_1)(M_1)+(1-N_1)(M_1)]\\\nonumber
&+&\langle\tilde{0_1}\tilde{0_2}|(q^{n_1})^{(\frac{1-x}{2})}(a_2)^{1-x}(q^{1-n_1})^{\frac{x}{2}}(a_1)^x[(1-M_1)(N_1)+(1-N_1)(1-M_1)].
\end{eqnarray}
In above equations $a,\;b\; \mbox{and}\; c$ are annihilation operators for states $|x\rangle,\;|y\rangle\;\mbox{and}\;|z\rangle$ respectively and summation runs over $x,\;y\; \mbox{and} \;z$ from 0 to 1.
\section{Conclusion}
The q deformation plays an essential role both in mathematics and physics. Although it is used to solve some complicated integrable systems it is applied to particle algebras. Since every force carrying particles obey boson algebra and other fundamental particles obey fermion algebra, deforming these algebras helps us understanding the behavior of them better. Nowadays, some studies show us q deformed systems better fits the experimental results than usual systems, for example q-deformed three dimensional harmonic oscillator.

With this motivation, we have constructed qubit states by using q-deformed boson algebra with arbitrary parameters $\psi$ and  quantum gates acted on q-deformed qubits as quantum operators. Then restrictions on parameters has been discovered. As q-deformed systems fit better experiments, quantum gates which are constructed with the help of q-deformed states may be more effective.

Also in forth chapter we have constructed q-deformed quantum gates as q-deformed oscillator states and have acted to q deformed qubits. Then to satisfy Eqs.(36-41) we have found arbitrary parameters which are used to write q-deformed gates and qubits and we found that these parameter $\psi$'s are different from non deformed quantum gate case.
As one can realize that in the third chapter non-deformed quantum gates act on q-deformed qubit, we defined arbitrary parameters $\psi$. In addition, in the forth chapter q-deformed quantum logic gates act on q-deformed qubit, arbitrary parameters has been defined and they are different form parameters which have been found in chapter three.

As a further study, we believe that studying q-deformed versions of higher level quantum systems and gates could lead to interesting findings.


\begin{thebibliography}{99}
\bibitem{drinfeld}Drinfeld, V. G.:Quantum Groups. Proceedings of International Congress of Mathematicians, Berkeley, California, USA.
(Academic Press, 1986), 798.
\bibitem{wor}Woronowicz, S. L.: Compact Matrix Preudogroups. Commun. Math. Phys. \textbf{111,} 613 (1987).
\bibitem{manin}Manin, Y. I.: Quantum groups and non-commmutative geometry (Centre De Recherches Math., Montreal,
1988).
\bibitem{ali} Arik, M. ,Baykal, A.: Inhomogeneous Quantum Groups for Particle Algebras. J. Math. Phys. \textbf{45,}  4207 (2004).

\bibitem{ali2} Altintas, A. A., Arik, M., and Atakishiyev, N. M.: On Unitary Transformations of Orthofermion Algebra that Form a Quantum Group. Mod. Phys. Lett. A \textbf{21} 1463 (2006).

\bibitem{ali3} Altintas, A. A. and Arik, M.: Quantum Groups of Fermionic Algebras Cent. Eur. J. Phys. \textbf{5,} 70 (2007).

\bibitem{ali4} Altintas, A. A., Arik, M.: Inhomogeneous Quantum Group Generalization of IO(2n,C) and ISp(2n,C):  Mod. Phys. Lett. A \textbf{23,} 617 (2008).

\bibitem{ali5} Altintas, A. A., Arik, M.: The Inhomogeneous Invariance Quantum Supergroup of Supersymmetry Algebra. Phys. Lett. A. \textbf{372,} 5955 (2008).

\bibitem{ali6} Altintas, A. A., Arik, M. and Arikan, A. S.: The Inhomogeneous Quantum Invariance Group of the Multi-Dimensional q-Deformed Boson Algebra. Cent. Eur. J. Phys. \textbf{8,} 131 (2010).

\bibitem{ali7} H. Alim, Altintas, A. A., Arik, M., Arikan, A. S.: The Inhomogeneous Qunatum Invariance Group of The Two Parameter
  Deformed Boson Algebra. Int. J. Theor. Phys. \textbf{49,} 633 (2010).
  
\bibitem{ali8} Altintas, A. A., Arik, M. and Arikan, A. S.: The Ingomogeneous Invariance Quantum Group of q-Deformed Boson Algebra with Continuous Paramters. J. Nonlinear Math. Phys. \textbf{18,} 121 (2011).

\bibitem{ali9} Altintas, A. A., Arik, M., Arikan, A. S. and Dil, E: Inhomogeneous Quantum Invariance Group of Multi-Dimensional Multi-parameter Deformed Boson Algebra. Chin. Phys. Lett. \textbf{29,} 010203 (2012).

\bibitem{ali10}Altintas, A. A.: Bosonic Algebras and Their Inhomogeneous Invariance Quantum Groups. 
J. Phys.: Conf. Ser. \textbf{343,} 012010 (2012).
\bibitem{arik}Arik, M., Coon, D. D.: Hilbert Spaces of Analytic Functions and Generalized Coherent States.  J. Math. Phys. \textbf{17,} 524 (1976).
\bibitem{arik2}Arik, M., Coon, D. D., Lam, Y .M.: Operator Algebra of Dual Resonance Model. J. of Math. Phys. \textbf{9,}  1776 (1975).
\bibitem{macfarlane}Macfarlane, A.J.: On q-Analogues of the Quantum Harmonic Oscillator and The Quantum Group $SU(2)_q$.
 J. Phys. A  \textbf{22,} 4581 (1989).
\bibitem{biedenharn}Biedenharn ,L. C.: The Quantum Group $SU_q(2)$ and a q-Analogue of The Boson Operators.  J. Phys. A \textbf{22,} L873 (1989).
\bibitem{bonatsos}Bonatsos, D.,Daskoloyamis, C.: Generalized Deformed Oscillators for Vibrational Spectra of Diatomic Molecules.
  Phys. Rev A \textbf{46,} 75 (1992).
\bibitem{bonatsos2} Bonatsos, D., Lewis, D., Raychev, P.P., Terziev, P.A.: Deformed Harmonic Oscillators for Metal Clusters:Analytic Properties and Supershells.  Phys. Rev. A \textbf{65,} 033203 (2002).
\bibitem{Georgieva}Georgieva, A. I., Sviratcheva, K. D., Ivanov, M. I., Draayer, J. P.: q-Deformation of Symplectic Dynamical Systems in ALgebraic Models of Nuclear Structure.  Phys. Atom. Nucl. \textbf{74,} 884, (2011).
\bibitem{Berrada} Berrada, K., El Baz, M., Eleuch, H., Hassouni, Y.: Bipartite Entanglement of Nonlinear Quantum Systems in The Context of The 
q-Heisenberg Weyl Algebra. Quant. Info. Proc. \textbf{11,} 351 (2012).
\bibitem{Berrada2}Berrada, K., Hassouni, Y.: Maximal Entanglement of Bipartite Spin States in The Context of Quantum Algebra.
 EPJD, \textbf{61,} 513 (2011).
\bibitem{Sun}Liu, Y. X., Sun, C.P., Yu, S. X., Zhou, D.L.: Semiconductor-Cavity QED in High-Q Regimes With q-Deformed Bosons. 
PRA \textbf{63,} 023802 (2001).

\bibitem{Sharma} Sharma, S.S.: Quantum Logic Gates with Ion Trap in an Optical Cavity. IJPMA \textbf{18,} 2221 (2003). 

\bibitem{Hasegawa} Hasegawa, H: Quantum Fisher Information and q-Deformed Relative Entropies. PTP \textbf{162,} 183 (2006).

\bibitem{SchwingerBook} J. Schwinger, Quantum Mechanics, Symbolism of Atomic Measurements, (Ed. by B-G. Englert, Springer, 2001).

\bibitem{RefNC} M.A. Nielsen and I.L. Chuang, Quantum Computation and Information. Cambridge University Press (Cambridge) (2000).

\bibitem{Gangopad2}Gagopadhyay, D.: The CNOT Quantum Logic Gate Uing q-Deformed Oscillators. Int. J. Quant. Infor., \textbf{06,} 471 (2008).

\bibitem{ToffoliFredkin1982} Fredkin, E., Toffoli, T.: Conservative Logic,  Int. J. Theor. Phys. \textbf{21,} 219 (1982).

\bibitem{BuguPRA2013A} Bugu, S., Yesilyurt, C., Ozaydin, F.: Enhancing the W-state quantum-network-fusion process with a single Fredkin gate, Phys. Rev. A \textbf{87,} 032331 (2013).

\bibitem{YesilyurtQInP2013A} Yesilyurt, C., Bugu S., Ozaydin, F.: An optical gate for simultaneous fusion of four photonics W or Bell states,  Quant. Info. Proc. \textbf{12,} 2965 (2013).

\bibitem{Ozaydin2014PRA}
Ozaydin, F., Bugu, S., Yesilyurt, C., Altintas, A. A., Tame, M., Ozdemir, S. K.: Fusing multiple W states simultaneously with a Fredkin gate, Phys. Rev. A  \textbf{89,} 042311 (2014).

\bibitem{lohe}Biedenharn, L.C., Lohe, M. A.  Quantum Group Symmetry
and q Tensor Algebras : World Scientific (Singapore) 1995.

\bibitem{Gangopad} Filippov, A.T., Gagopadhyay, D., Isaev, A.P.: Harmonic Oscillator Realization of The Canoncical q-Transformation. J.Phys. A \textbf{24,} L63 (1991).

\end{thebibliography}
\end{document}